\documentclass[twocolumn,preprintnumbers,amsmath,amssymb,superscriptaddress]{revtex4}

\usepackage[dvips]{graphicx}
\usepackage{dcolumn}
\usepackage{bm}
\usepackage{xspace}
\usepackage{multirow}
\usepackage{natbib}
\usepackage{color}

\begin{document}

\preprint{}
\title{Reconstruction of the Bi$_2$Sr$_2$CaCu$_2$O$_{8+\delta}$ Fermi Surface}
\author{T. Valla}
\affiliation{Condensed Matter Physics and Materials Science Department, Brookhaven National Lab, Upton, New York 11973, USA\\}
\email{valla@bnl.gov}
\author{I. Pletikosi\'{c}}
\affiliation{Condensed Matter Physics and Materials Science Department, Brookhaven National Lab, Upton, New York 11973, USA\\}
\affiliation{Department of Physics, Princeton University, Princeton, NJ 08544, USA}
\author{I. K. Drozdov}
\affiliation{Condensed Matter Physics and Materials Science Department, Brookhaven National Lab, Upton, New York 11973, USA\\}
\author{G. D. Gu}
\affiliation{Condensed Matter Physics and Materials Science Department, Brookhaven National Lab, Upton, New York 11973, USA\\}

\date{\today}

\begin{abstract}
The effects of structural supermodulation with the period $\lambda \approx26$ \AA \ along the $b$-axis of Bi$_2$Sr$_2$CaCu$_2$O$_{8+\delta}$ have been observed in photoemission studies from the early days as the presence of “diffraction replicas” of the intrinsic electronic structure. Although predicted to affect the electronic structure of the Cu-O plane, the influence of supermodulation potential on Cu-O electrons has never been observed in photoemission. In the present study, we clearly see, for the first time, the effects on the Bi$_2$Sr$_2$CaCu$_2$O$_{8+\delta}$ electronic structure - we observe a hybridization of the intrinsic bands with the supermodulation replica bands in the form of avoided crossings and a corresponding reconstruction of the Fermi surface. We estimate the hybridization gap, $2\Delta_h\sim25$ meV in the slightly underdoped samples. The hybridization weakens with doping and the anti-crossing can no longer be resolved in strongly overdoped samples. In contrast, the \lq\lq{}shadow\rq\rq{} replica, shifted by $(\pi, \pi)$, is found not to hybridize with the original bands within our detection limits. 
\end{abstract}
\vspace{1.0cm}

\pacs {74.25.Kc, 71.18.+y, 74.10.+v, 74.72.Hs}

\maketitle

By interfacing layers of two 2D materials with slightly different lattice constants, a new, long wavelength superstructure (moir\'{e} pattern) appears. \cite{Pletikosic2009,Wilson2017} The similar effect occurs when layers of the same material are twisted relative to one another \cite{LopesdosSantos2007,Bistritzer2011,Yao2018} Depending on the coupling between the layers and the strength of the moir\'{e} superlattice potential, the electronic structure of the system will be affected and could lead to novel phenomena at some "magic" mismatches or twist angles. \cite{Cao2018,Cao2018a,Chen2019} The most direct probe of the electronic structure, the Angle-Resolved Photoemission Spectroscopy (ARPES), where a photon is absorbed and an electron is emitted, provides not only an intrinsic information about the system\rq{}s electronic structure, but also about the outgoing electron's path on the way to detector. In the case where the supermodulation does not affect the electronic structure of constituent layers, replica bands could still appear due to diffraction of outgoing electrons on a supermodulation. In that case, there will be no hybridization between the original bands and replicas. \cite{Norman1995} If there is a finite influence of the supermodulation, the original and replica bands should hybridize and the gaps should appear at the points of crossings. 

Bi$_2$Sr$_2$CaCu$_2$O$_{8+\delta}$ (Bi2212) is a prototypical cuprate high-Tc superconductor (HTSC) and possibly one of the most studied materials by ARPES that led to significant contributions to our understanding of cuprates. The $d$-wave symmetry of the superconducting (SC) gap, \cite{Shen1993,Damascelli2003} the normal state gap (pseudogap), \cite{Ding1996,Marshall1996} the quasiparticle (QP) self-energy \cite{Valla1999,Johnson2001a,Damascelli2003,Valla2007,Valla2019} are a few notable examples. However, there are still many remaining questions, including the one related to the role of well-known crystallographic superstructure on the electronic properties of Bi2212. \cite{Norman1995,Ding1996b}
In Bi2212 crystals, the mismatch between the Bi-O and Cu-O planes causes a superstructure with the period $\lambda \approx26$ \AA, running along the crystallographic $b$ direction. \cite{Withers1988} The effect of this supermodulation has been observed in ARPES very early on as replicas of the original bands. \cite{Norman1995} From the lack of any detectable hybridization between the original and replica bands, it was concluded that the effect of supermodulation on the electronic structure of Cu-O planes is zero and that replicas represent electrons from undisturbed Cu-O planes, diffracted on the supermodulated Bi-O planes on their way to vacuum.

Scanning Tunneling Microscopy (STM) experiments, however, have indicated that the pairing potential inside the Cu-O planes is actually affected by this supermodulation - the superconducting gap has been shown to be modulated in the $b$-direction with the same period and the phase as the structural supermodualtion. \cite{Andersen2007,Slezak2008} These results would imply that the superlattice actually alters the intrinsic electronic structure of Cu-O planes. Therefore, a closer look at the electronic structure at the regions of reciprocal space where the replicas cross the original Fermi surface (FS) (in the X quadrant of the Brillouin zone) would be extremely important. 

In the present ARPES study, we clearly detect, for the first time, the effects of supermodulation on the Bi2212 electronic structure. We observe the hybridization of the original bands with the replica bands in the form of avoided crossings and an accompanying reconstruction of the FS. We expect that the reconstruction should affect the low energy properties of this material, not only in the superconducting, but also in the normal state near $T_c$ where the states remain sufficiently coherent and should be observable by other probes. We also find that the "shadow" replica, shifted by $(\pi, \pi)$, does not hybridize with the original bands. 

\begin{figure}[htpb]
\begin{center}
\includegraphics[width=8cm]{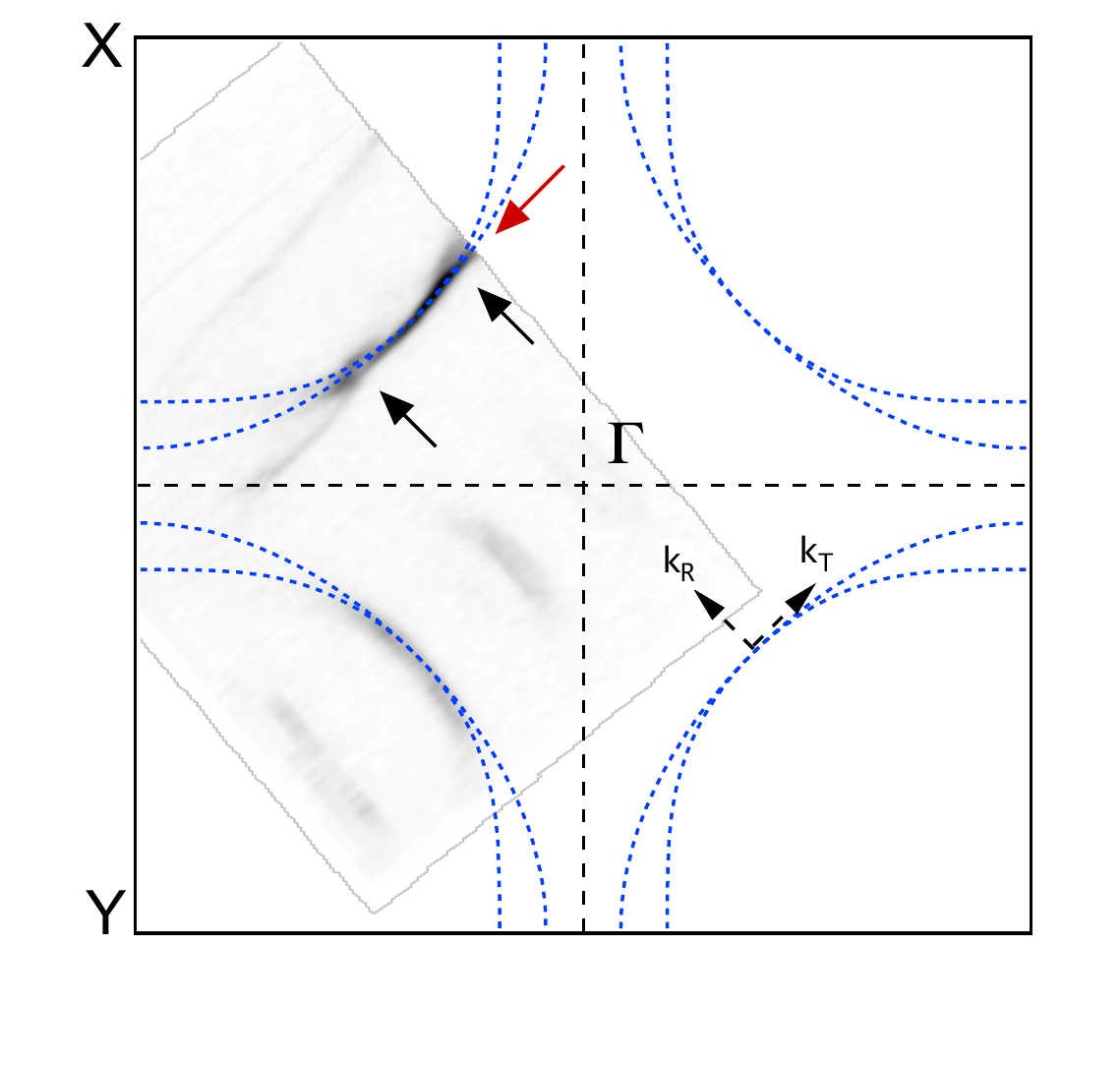}
\caption{Reconstructed Fermi surface of Bi$_2$Sr$_2$CaCu$_2$O$_{8+\delta}$. Constant energy contour at $E=-5$ meV of ARPES intensity from slightly underdoped (UD87) Bi2212, taken in the superconducting state (T=25 K). The black arrows indicate the reconstruction due to the structural supermodulation in the form of avoided crossings between the original FS and supermodulation replicas. The red arrow indicate the quasi-1D segments formed by reconstruction. The dashed arrows indicate the coordinate axis used in Fig. \ref{Fig2}, with the origin at the nodal point. The map was recorded with the analyzer slit in the $k_R$ direction, by rotating the sample in the $k_T$ direction in steps of 0.5$^{\circ}$. The blue curves represent the tight-binding fit of the experimental FS. 
}
\label{Fig1}
\end{center}
\end{figure}

The experiments within this study were done in an experimental facility that integrates oxide-molecular beam epitaxy with ARPES and STM capabilities within the common ultra high vacuum system. \cite{Kim2018a} The starting sample was a slightly overdoped Bi2212 with $T_c=91$ K (OD91), synthesized by the traveling floating zone method, clamped to the sample holder and cleaved with Kapton tape in the ARPES preparation chamber. The silver-epoxy glue, commonly used for mounting samples, was completely eliminated, resulting in large and perfectly flat cleaved surfaces, critical for the observation of small effects reported here. Reduction in doping was achieved by subsequent annealing of the same surface in vacuum to 350$^{\circ}$C, resulting in $T_c=87$ K (UD87). 
The ARPES experiments were carried out on a Scienta SES-R4000 electron spectrometer with the monochromatized HeI (21.22 eV) radiation (VUV-5k). The total instrumental energy resolution was $\sim$ 5 meV. Angular resolution was better than $\sim 0.1^{\circ}$ and $0.3^{\circ}$ along and perpendicular to the slit of the analyzer, respectively. 

\begin{figure}[htpb]
\begin{center}
\includegraphics[width=7.5cm]{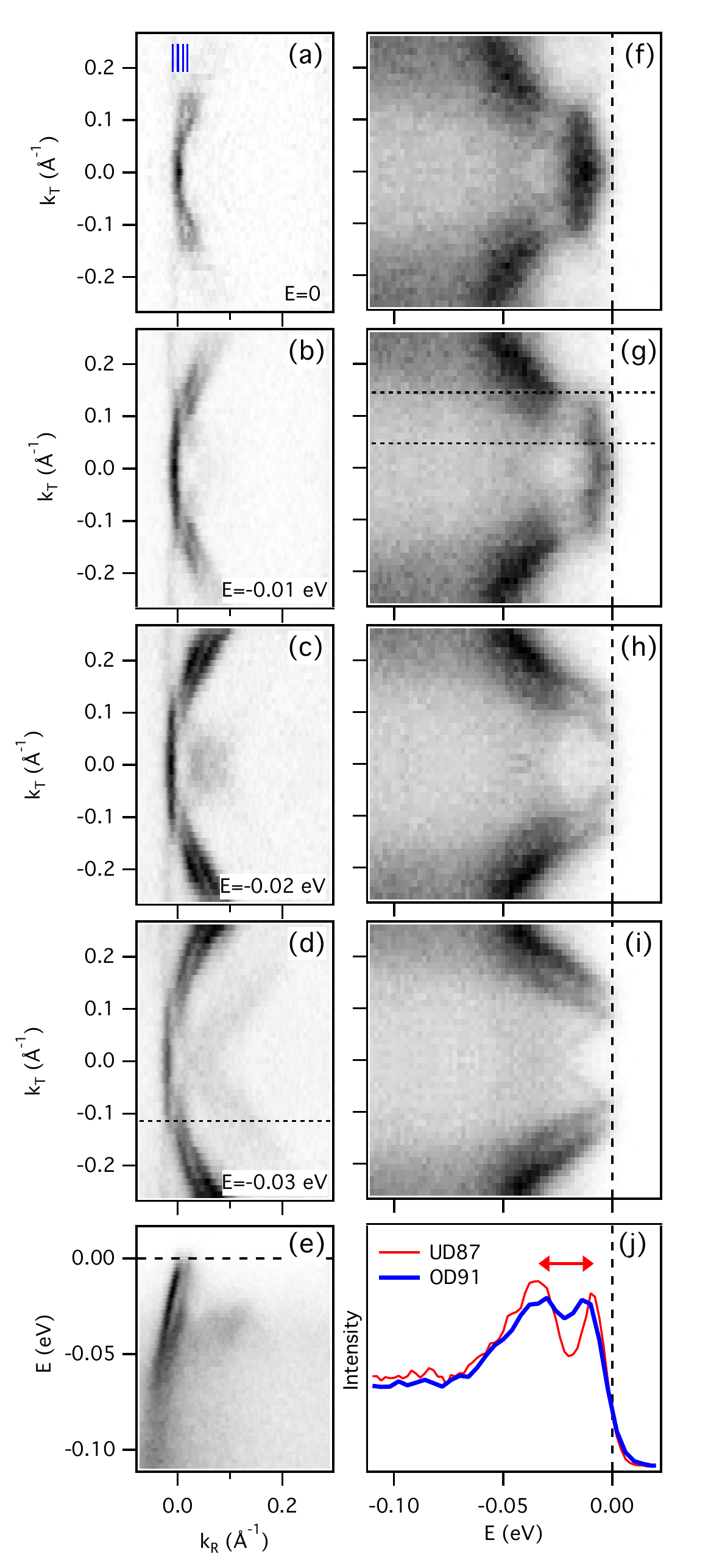}
\caption{Reconstructed electronic structure of Bi2212. (a-d) Constant energy contours of ARPES intensity near the nodal region in the $\Gamma-X$ quadrant of the Brillouin zone at energies as indicated. (e) The ARPES spectrum taken along the momentum line indicated in panel (d). (f-i) The ARPES spectra taken along the four momentum lines nearly tangential to the nodal segment of the FS as indicated in panel (a). The top (bottom) spectrum corresponds to the left (right) momentum line in (a). The momenta are measured from the nodal Fermi wave vector. All the spectra were taken with the analyzer slit in the $k_R$ direction, while the mapping in the $k_T$ direction was done by rotating the sample in steps of $1/3$ of a degree. (j) The energy distribution curve integrated over the momentum region between the two dotted lines in panel (g) for UD87 and OD91. The arrow indicates the magnitude of the hybridization gap in the UD87 sample, $2\Delta_h\approx25$ meV.
}
\label{Fig2}
\end{center}
\end{figure}

Fig. \ref{Fig1} shows the low-energy ($E=-5$ meV) contour of the electronic structure of a slightly underdoped sample, UD87, as a function of of a wave-vector in the first Brillouin zone. Aside from the near-nodal segments of the original FS, the equally spaced replicas, running in the $\Gamma-\mathrm{Y}$ direction are clearly visible. They do not intersect in the Y quadrant, but in the X quadrant they do, at the points marked by black arrows. Already at this scale, the loss of intensity at the points of intersection is clearly visible, indicating avoided crossings and the reconstruction of the FS due to the impact of supermodulation potential on the Cu-O states. The resulting low-energy part of reconstructed FS takes the form of a wavy, quasi-1D segments, running in the $\Gamma-\mathrm{Y}$ direction, as indicated by the red arrows.

Fig. \ref{Fig2} shows the reconstructed electronic structure in more details. The panels (a-d) show the ARPES intensity in the near-nodal region of the FS for several energies, as indicated. The reconstruction is visible at energies $E>-50$ meV, while outside of that range, the lifetime broadening is too big for the anti-crossing to be resolved. As the constant energy contour becomes larger for energies away from the Fermi level, the anti-crossing stays at the same $k_T=\pi/\lambda\approx\pm0.121$ \AA$^{-1}$, effective zone boundaries created by the supermodulation potential. \cite{Voit2000} The panel (e) shows the dispersion of the states along that momentum line, while panels (f-i) show the dispersion along the four momentum lines nearly tangential to the nodal Fermi segments, as marked in panel (a). The hybridization gap can be easily estimated from spectra like those in panels (f-h), $2\Delta_h\approx25$ meV, for the UD87 sample and $\approx20$ meV for the as grown, OD91 sample. We have also studied more overdoped samples where the reconstruction could no longer be detected. \cite{Drozdov2018} 

As can be seen in Fig. \ref{Fig2}, the reconstruction occurs at the points where the bilayer splitting is already sizable and our results suggest that it affects both the bonding and the antibonding states. We also note that the hybridization can be observed in the normal state, relatively close to $T_c$ and only in the slightly underdoped samples. This is not surprising, as at higher temperatures the lifetime broadening becomes larger than $\Delta_h$ and the reconstruction can no longer be resolved. The finding that $\Delta_h$ decreases with doping would suggest that the superstructure potential becomes better screened as more carriers are doped to the Cu-O planes.

Also visible from Fig. \ref{Fig2} is an incomplete gapping of spectral intensity at avoided crossings. This is obvious in constant energy contours (a-d), dispersion plots (e-i), as well as in the EDCs (j). In addition, we observe that the intensity of replicas does not decay away from the anti-crossings (Fig. \ref{Fig1}) as it should if the hybridization alone was at play. In the mean field like picture intensities should decay in a BCS-like manner. \cite{Norman1995} Therefore, it appears that the total intensity of replicas is a combination of hybridization and diffraction of the outgoing Cu-O electrons on the Bi-O supermodulation. That would also explain the incomplete gapping. Within this picture, the intrinsic 1D segment of the reconstructed FS decay rapidly away from the anti-crossings, while the diffraction extends the spectral weight uniformly over the whole diffraction order.

\begin{figure}[htpb]
\begin{center}
\includegraphics[width=7.5cm]{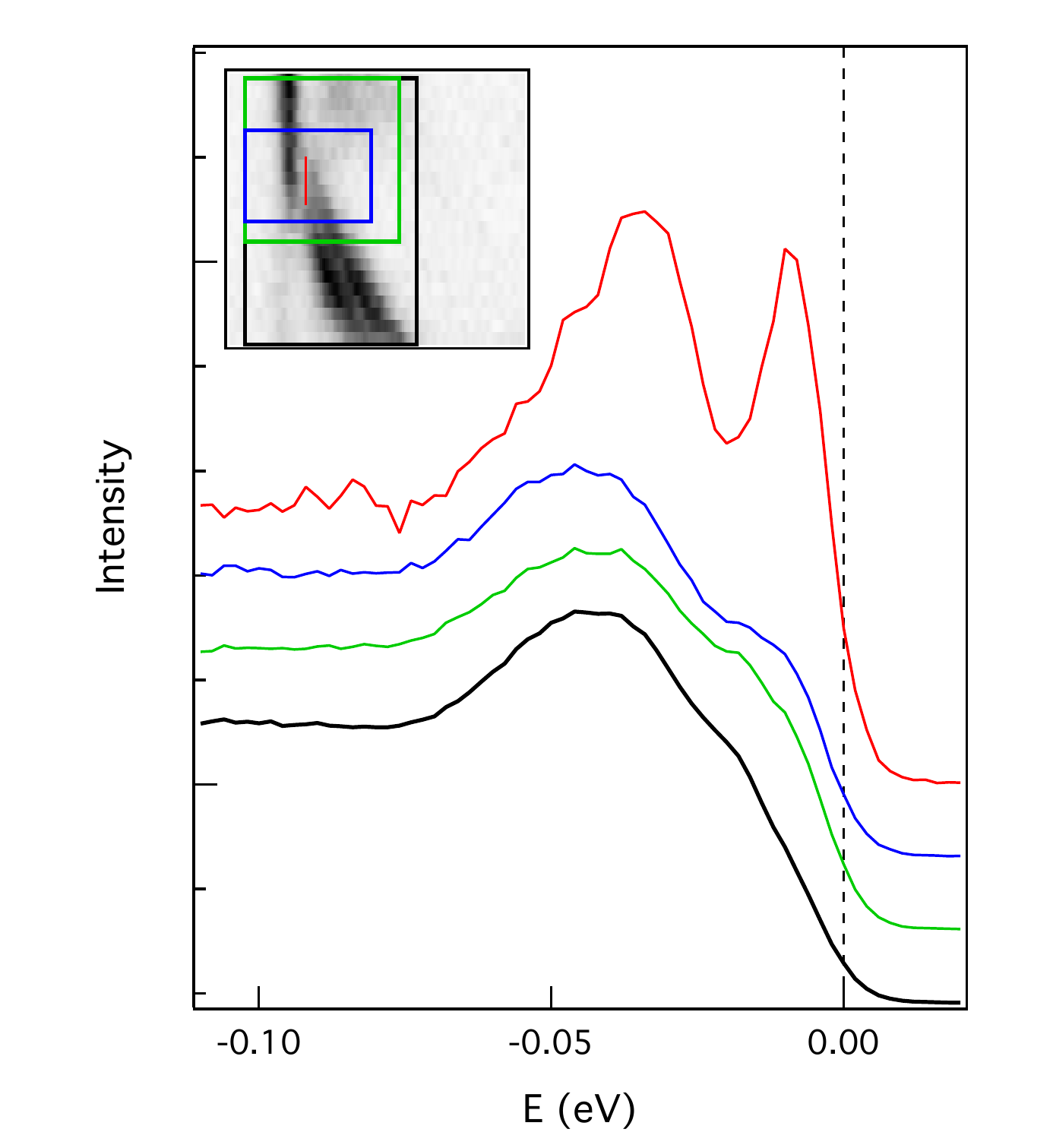}
\caption{Hybridization gap and density of states. The energy distribution curve from Fig.2(j) (red) and several other EDCs obtained by integrating the spectral intensity over the $k$-regions as indicated in the inset, with the corresponding colors. 
}
\label{Fig3}
\end{center}
\end{figure}

Could this effect be observed by other probes? Aside from the mentioned modulation of the superconducting gap, we think that there are additional indications of the hybridization gap in the published  STM data. For example, the local density of states (LDOS) in the superconducting state often shows a prominent shoulder at energies well inside the maximal gap, at $\approx$ 20-25 meV. \cite{Pan2001} We believe that this originates from the contribution to LDOS from the hybridization gap. To illustrate this, we show in Fig. \ref{Fig3} the energy distribution curves (EDC) where the electronic states are integrated over several different $k$-regions, as indicated in the inset. It is clear that, in the superconducting state, the low-energy shoulder in the EDCs is always present, but its contribution weakens as we increase the integration box and approach the total DOS. Note that the black integration box already includes a significant portion of the Brillouin zone and that its further expansion will not additionally affect the DOS near 20-25 meV. The unaffected node from the Y quadrant will contribute a regular $d$-wave DOS in the equal amount, but the structure from the X quadrant will remain in the total DOS. This demonstrates that the DOS inevitably contains a contribution from hybridization and calls for a re-examination of possible causes for the modulation of gaps observed in STM. \cite{Slezak2008} Obviously, if the electronic structure is modulated along the $b$ axis, so would have to be the superconductivity, as initially speculated by Norman \textit{et al}. \cite{Norman1995} We also note that a significant in-plane anisotropy has been observed in DC transport and in optical conductivities, both in the normal and superconducting states. \cite{Quijada1994,Quijada1999} Our results offer a straightforward explanation for these anisotropies.

\begin{figure}[htpb]
\begin{center}
\includegraphics[width=9.2cm]{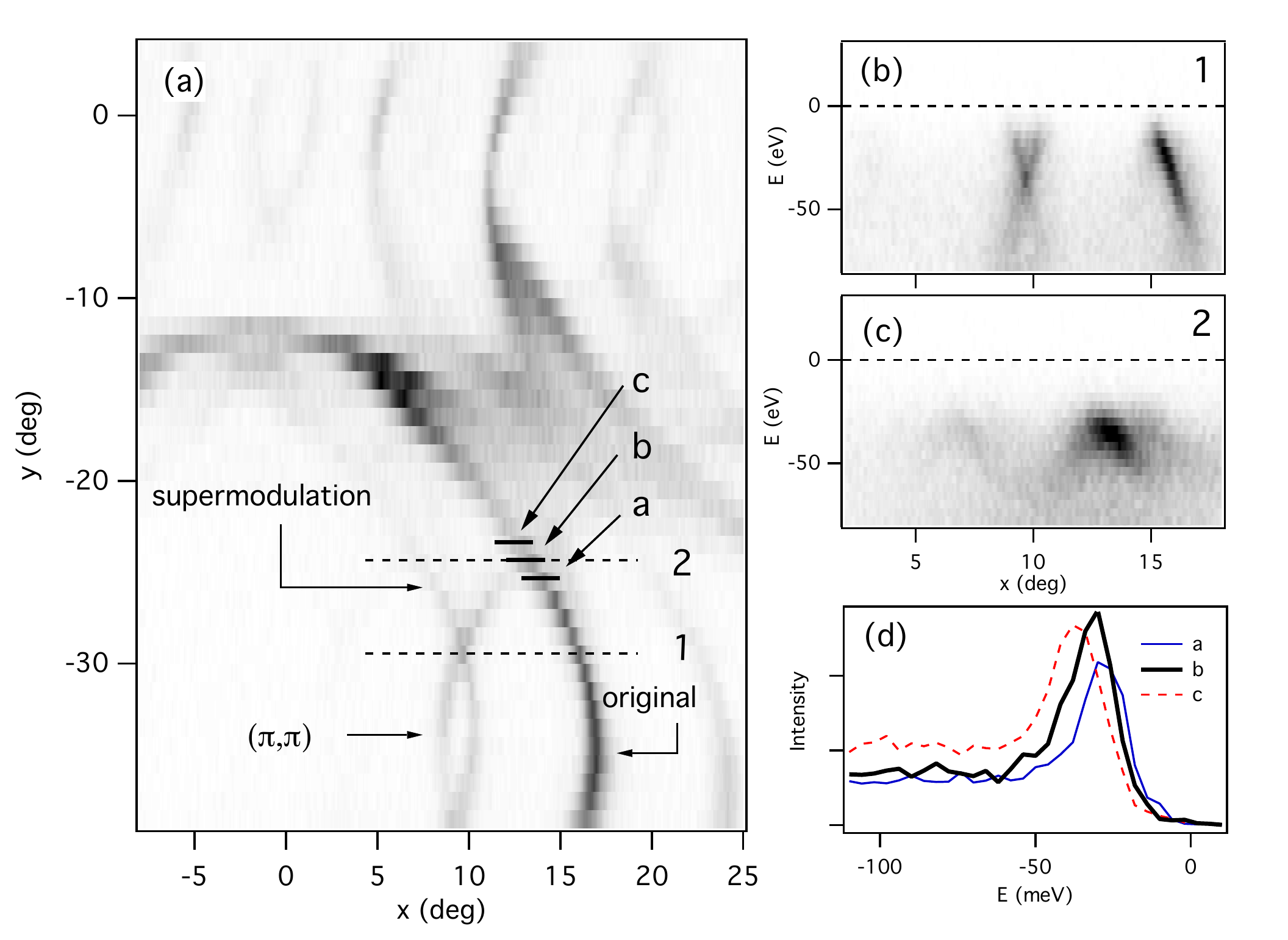}
\caption{The $(\pi, \pi)$ replica. (a) The constant energy contour of ARPES intensity from OD91 sample recorded at $E=-26$ meV in the superconducting state ($T=25$ K). (b,c) ARPES spectra showing the dispersion of the states along the lines indicated in (a). The $(\pi, \pi)$ replica does not hybridize with the supermodulation replica (line 1) or with the original band (line 2). The raw intensities are shown as functions of the emission angle. The mapping was done with the analyzer slit in the $x$ direction, by rotating the sample in the $y$ direction in steps of 1$^{\circ}$ (d) EDCs taken at the points on the original FS as indicated in (a).
}
\label{Fig4}
\end{center}
\end{figure}

Aside from the replicas originating from the $\approx26$ \AA\ supermodulation, it has been known from the early days that ARPES studies show another replica of the original electronic structure - the one that is shifted by the $(\pi, \pi)$ wavevector relative to the original. \cite{Ding1996b,Mans2006} By realizing that the $\approx26$ \AA\ supermodulation affects the Cu-O planes, it is obviously interesting to check if the the $(\pi, \pi)$ superstructure has a similar effect. In Fig. \ref{Fig4}, we show that the $(\pi, \pi)$ replica behaves differently in that there is no observable hybridization with the original bands, nor with the supermodulation replicas. The avoided crossings occur neither in the constant energy contours, Fig. \ref{Fig4}(a), nor in the dispersions, Fig. \ref{Fig4}(b,c). Furthermore, the EDCs taken at the crossing point of the original FS with the $(\pi, \pi)$ replica, and just away from it, show a single coherence peak with no sign of a splitting or a change in the spectral width, putting the upper limit on the hybridization gap to $2\Delta_h\lesssim4$ meV. This suggests that the $(\pi, \pi)$ replica is most likely the diffraction image of outgoing photoelectrons on the superstructure in the layers, not affecting the Cu-O planes in any significant way. We note that Bi2212 has an orthorhombic unit cell, which by itself could produce doubling of the bands and generate the $(\pi,\pi)$ replica. \cite{Mans2006} 

In conclusion, we have detected modulation of the electronic structure of Bi2212 due to the $\approx26$ \AA\ structural supermodulation, resulting in the first direct observation of reconstruction of the Bi2212 FS. In contrast, the $(\pi, \pi)$ replica does not hybridize with the original bands of the Cu-O planes. We also note that a number of different phenomena have been proposed to occur in the underdoped cuprates that should also result in the reconstructed electronic structure. \cite{Doiron-Leyraud2007,Yang2008,Yang2011,Ghiringhelli2012,Fujita2014,Badoux2016,Norman2018} These reconstructions, although expected to be quite significant and easily detectable with the current ARPES resolution, remain elusive, \cite{Drozdov2018} with the exception of the one seen at the anti-ferromagnetic zone boundary in the electron-doped cuprates. \cite{Armitage2002,He2019} 

\section*{Acknowledgements}
This work was supported by the US Department of Energy, Office of Basic Energy Sciences, contract no. DE-SC0012704. I.P. is supported by  ARO MURI program, grant W911NF-12-1-0461.

\bibliographystyle{apsrev}
\bibliography{Cuprates,Reconstruction}

\end{document}